\documentclass{llncs}
\usepackage{times}
\usepackage{balance}
\usepackage{url}
\usepackage{epsfig}
\usepackage{graphicx}
\usepackage{subfigure}
\usepackage{amsmath}
\usepackage{amssymb} 
\usepackage{amsfonts}
\usepackage{color}
\usepackage{multirow}
\usepackage{algorithm}
\usepackage[utf8]{inputenc}
\usepackage[T1]{fontenc}

\newcommand{\para}[1]{{\vspace{1pt} \bf \noindent #1 \hspace{10pt}}}

\begin{document}
\title{On the Bursty Evolution of Online Social Networks}

\author{Sabrina Gaito$^\dag$, Matteo Zignani$^\dag$, Gian Paolo Rossi$^\dag$, 
   Alessandra Sala$^\S$\\ Xiaohan Zhao$^\ddag$, Xiao Wang$^*$,
  Haitao Zheng$^\ddag$, Ben Y. Zhao$^\ddag$\\
  $^\dag${\tt Department of Computer Science, Universit\'a degli Studi di Milano}\\
  $^\S${\tt Bell Labs, Ireland}  \hspace{0.5in}    $^*${\tt Renren Inc.}\\
  $^\ddag${\tt Computer Science, U. C. Santa Barbara} 
}

\maketitle

\begin{abstract}
  The high level of dynamics in today's online social networks (OSNs) creates
  new challenges for their infrastructures and providers. In particular,
  dynamics involving edge creation has direct implications on strategies for
  resource allocation, data partitioning and replication.  Understanding 
  network dynamics in the context of physical time is a critical first step
  towards a predictive approach towards infrastructure management in OSNs.
  Despite increasing efforts to study social network dynamics, current
  analyses mainly focus on change over time of static metrics computed on
  snapshots of social graphs.  The limited prior work
  models network dynamics with respect to a logical clock. In this paper, we
  present results of analyzing a large timestamped dataset describing the
  initial growth and evolution of Renren, the leading social network in
  China. We analyze and model the burstiness of link creation process, using
  the second derivative, {\em i.e.} the acceleration of the degree. This
  allows us to detect bursts, and to characterize the social activity of a
  OSN user as one of four phases: acceleration at the beginning of an
  activity burst, where link creation rate is increasing; deceleration
  when burst is ending and link creation process is slowing;
  cruising, when node activity is in a steady state, and complete
  inactivity.
\end{abstract}

\keywords{online social networks, network dynamics}

\section{Introduction}


The rapid growth of online social networks (OSNs) has created numerous technical
challenges for the providers that supply the hardware and software
infrastructure behind these web services.  As one example, the creation of
social links between users dramatically change demands on social network
infrastructures in terms of access, storage and computation.  Depending on
the specific configuration of backend servers, for example, changes in the
social graph can affect how data is partitioned across clusters, or how much
replication is necessary to sustain low query response times.


However, very little is known about how social network dynamics
correspond to actual clock time.  The large majority of prior work on OSN
analysis has focused on analyzing, mining, and modeling static topologies or
static snapshots of dynamic processes.  Only recently have researchers begun
to study dynamic processes in social networks, most often by analyzing how
classical graph metrics such as degree, connected components, and shortest
paths change over time.  This has led to models of underlying processes such
as densification and shrinking diameters~\cite{leskovec2005graphs}.  These
models describe how graphs change and how edges are created with respect to a
logical clock, {\em i.e.}  a homogeneous sequence of events.


But how do these events match up to events in real time?  Can we better
understand how edge creation events relate to each other, and can the
occurrence of such events be predicted with respect to a physical clock?  This
work is an initial effort to 
answer some of these questions, but analyzing one specific temporal property
of {\em burstiness} in edge creation.  Our work is motivated in part by
models of human dynamics adopted in a wide range of disciplines, from
economics to communications.  Recent studies~\cite{Vazquez06,Kleinberg_burst} have shown that human dynamics are best described by
periods of rapidly occurring events interleaved with long periods of
inactivity.  Thus we ask the question: {\em Is link creation in online social
  networks a bursty process?} 

In this paper, we provide an initial answer to this question, by analyzing an
anonymized temporal trace of edge creation events over a period of a year in
a large OSN.  The trace describes events in Renren~\cite{renren-imc10}, the
largest social network in China with more than 220 million users. Our
analysis shows that edge creation is a highly inhomogeneous and bursty
process.  We then ask two followup questions: a) {\em Given an high level
  bursty structure, does an inner substructure exist, and how can it be
  characterized}; and b) {\em How can we detect both the whole burst and its
  internal phases?}


Understanding the internal structure of edge creation bursts can shed light
on the underlying user process, {\em e.g.} is the user gradually enlarging her circle
of friends or has she discovered a new cluster of her offline friends.  
Known techniques for the analysis and the detection of burst events
(gamma-ray, text mining, stock market) focus on locating a burst when it
occurs, but they do not consider events inside the detected temporal window.
Thus we propose a new methodology able to detect bursts, their internal
structure and the transitions between the different phases a node
experiences. We perform a second order analysis on the link creation process
by computing, for each node, the acceleration of the degree time
function to characterize the burst structure.

Finally, we apply our acceleration metric and the detection of bursty phases
on Renren.  We find that all nodes exhibit similar patterns over time,
characterized by an intense burst of activity following their joining the
network.  The initial burst is followed by weaker bursts over time, each
composed of an acceleration phase, followed by a longer period of slowly
vanishing deceleration.

The discovery of highly bursty patterns paves the way for new generative
models that not only capture graph dynamics in terms of phases of node
activity, but also describes such events with respect to physical time.  In
addition, burst analysis can reveal further insights into the formation and
liveness of individual users, communities, and provide a basic metric of
useful in characterizing and comparing different traces of network dynamics.


\section{Related Work}

\para{Time evolving OSN Snapshots.}
While static features of OSNs are well studied, works on dynamics of online
social networks are still ongoing. Among all Leskovec {\em et al.}
in~\cite{leskovec2005graphs} detected two important properties on dynamic OSN
data: graph densification, {\em i.e.} the average degree increases, and
shrinking diameter.  Several different social graphs has been studied in
order to capture the growth of components and communities. Palla {\em et al.}
\cite{Palla2007} investigated the time dependence of overlapping communities
and Berlingherio {\em et al.} \cite{Berlingherio2010} detected clusters of
temporal snapshots of a network, interpreted as eras of evolution. Authors
in~\cite{mcglohon2008weighted,zheleva2009co,akoglu2009rtg} studied the
dynamics of disconnected components. Finally Backstrom {\em et al.}~\cite{Backstrom2006} investigated the structural features which influence
people in joining communities and their growth process. Alternatively, the
per node dynamics was studied in~\cite{leskovec2008microscopic} where the
authors captured the evolution of key network parameters, and evaluate the
extent to which the edge destination selection process subscribes to
preferential attachment.  As concerns acceleration, in~\cite{dorogovtsev} the
authors considered an overall network size growth as a global property and
they modeled this global acceleration for the purpose of predicting the next
network stage.


\para{Interdisciplinary Study of Human Dynamics.}
In~\cite{barabasi:nature2005}, Barabasi observed that the timing of human
activity is inhomogeneous and bursty, disputing the previous hypothesis that
human activities are randomly distributed in time.  The inhomogeneity idea
was extended in \cite{Vazquez06} and validated on few networks such as an
Hungarian news portal, e-mails, library activities in an University and a
trade transactions.  Similarly, \cite{McGlohon07} analyzes the activity
burstiness of blogs using entropy plots, and show non-uniformity and
self-similarity of the number of posts time sequences.  Furthermore,
\cite{kumar11} and \cite{Eom11} observe that temporal patterns are inhomogeneous or bursty
even in mobile phone calls and in citation dynamics.
Finally, \cite{Kleinberg_burst} developed a burst detection algorithm and observed that the appearance of a topic in a stream of documents, such as e-mails or research papers, has a bursty behavior. 
As far as we know, burstiness has never been investigated in OSN.

\section{Timestamped Renren Dataset} 
A major obstacle to studying OSN dynamics is the difficulty of obtaining
detailed data traces.  Our study uses an anonymized dataset that contains the
timestamped creation of all users and edges in the Renren social
network~\cite{renren-imc10}.  Renren is the largest and oldest OSN in China,
with functionality similar to Facebook, and has currently over 220 million
users.  Our anonymized trace describes the evolution of each nodes as a
sequence of timestamped edge creation events.  The first edge created in the
Renren network dates back to November 2005.  We track the complete evolution
of the oldest 60000 nodes in Renren, for a total of 8 million edges created
over a period of one year from November 2005 to December 2006.

%
%
%

\section{bursty nature of link creation}

\label{sec4:intro}

\begin{figure}
\centering
\includegraphics[width=3.3in]{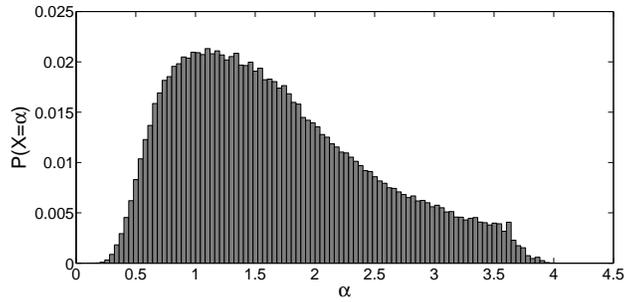}
\caption{Distribution of the scale parameter $\alpha$, which characterizes the
  inter-event time distribution between consecutive link creations for a
  single individual. $\alpha$ values have been grouped in bins of length
  0.05. Values past the peak around 1 decreases much more slowly
  with respect to the left side.}
\label{alphaDistro}
\end{figure}

\begin{figure*}
\centering
\subfigure[]{
	\includegraphics[width=3.3in]{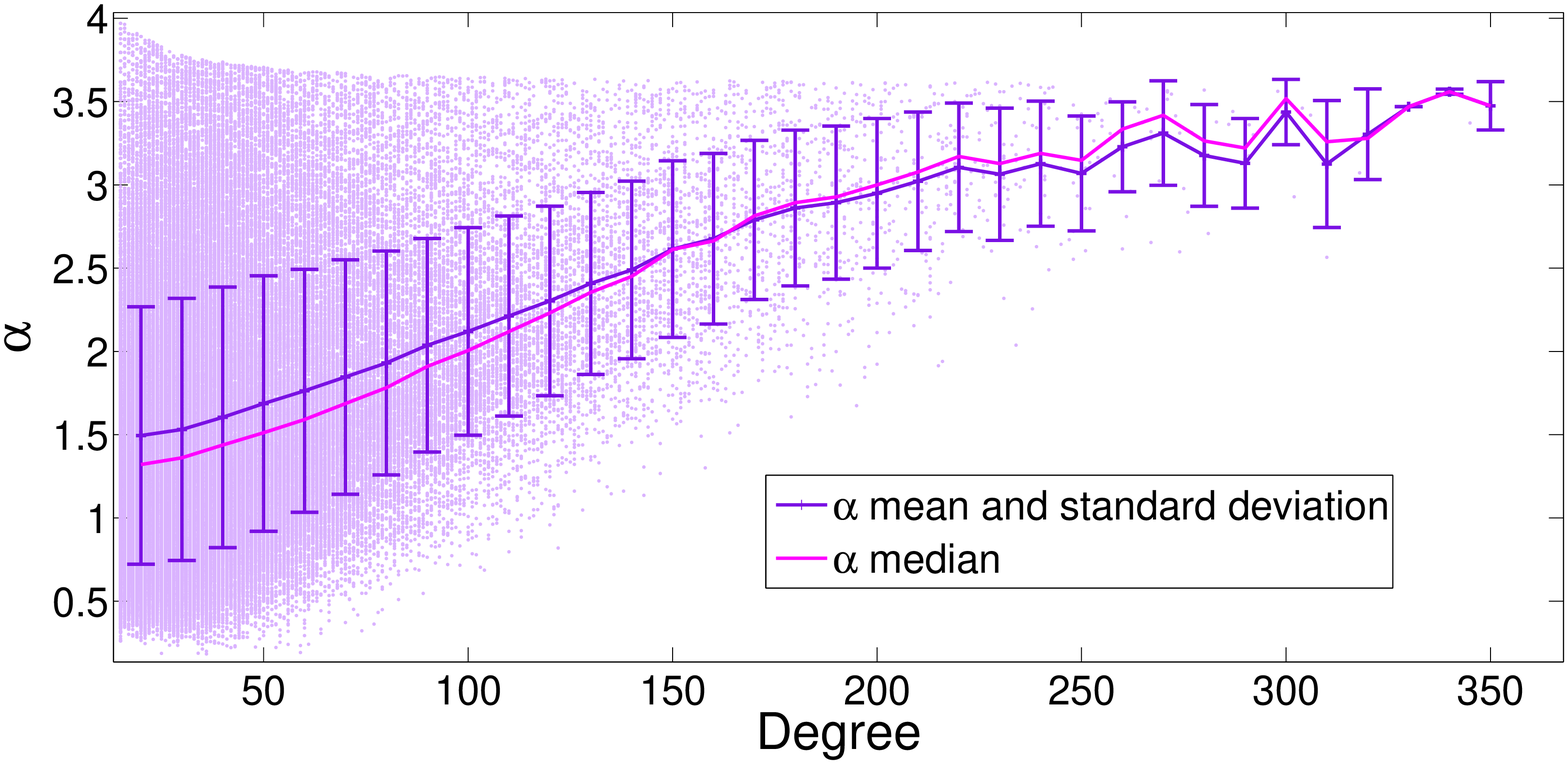}\label{alphaOnAge}
}
\subfigure[]{
	\includegraphics[width=3.3in]{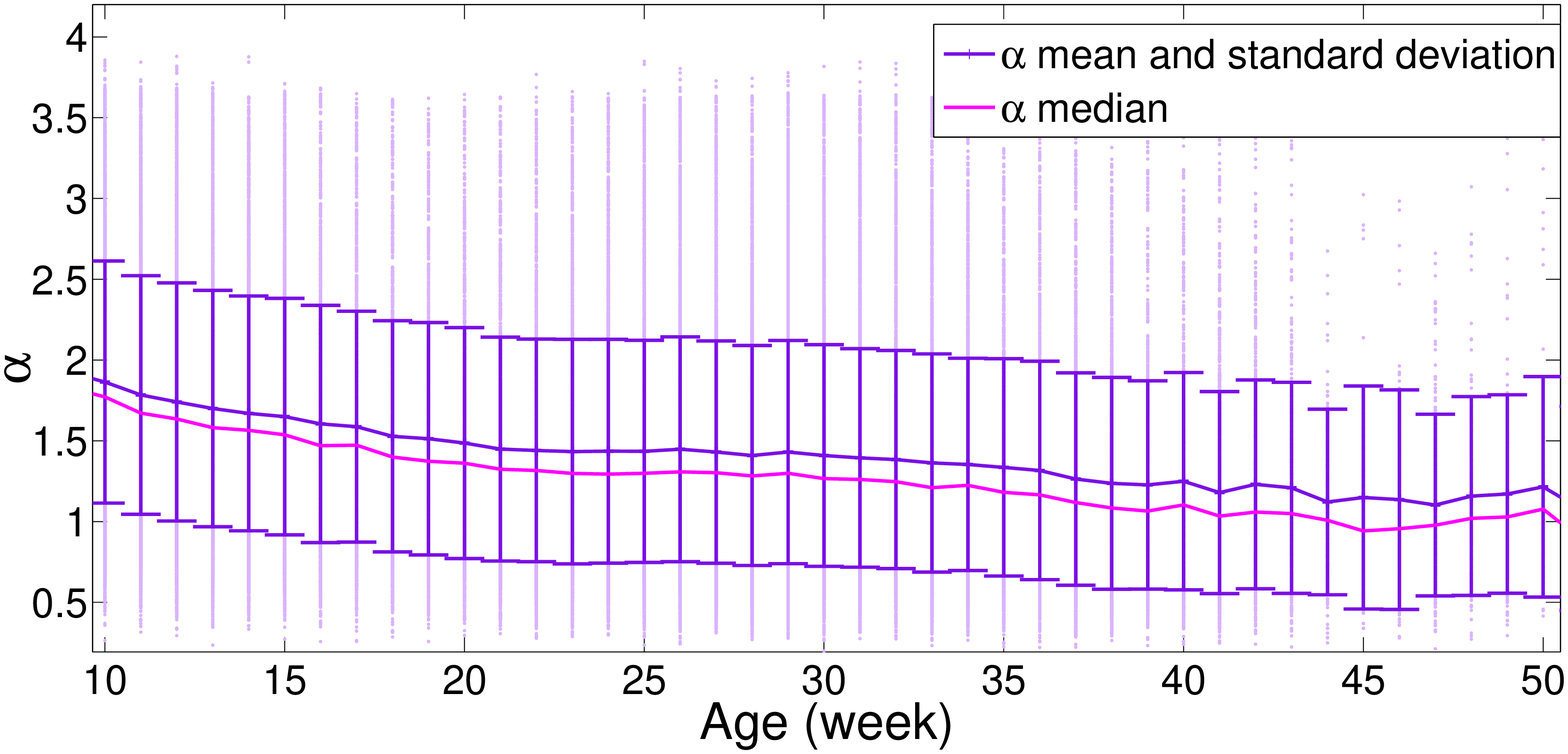}\label{alphaOnDegree}
}
\caption{ a) mean, median, and standard deviation (error bar) of $\alpha$ as
  function of the final degree. Mean and median increase with the final node
  degree.  To compute these values, we group node degrees in bins of 10, and
  consider their relative $\alpha$ values. b) mean, median and standard
  deviation (error bar) of $\alpha$ as function of node age measured in
  weeks.  Mean and median decrease very slowly with node age.}
\label{relations}
\end{figure*}

Bursty behavior has been observed in various contexts such as WWW traffic
patterns~\cite{Crovella97}, emails exchanges, and in general human
behavior~\cite{Vazquez06}.  But they have not been studied in the context
of online social networks.  In this section, we study the link creation
process as the growth of the neighborhood of each single node, and
show that the linking activity of online social networks users is
characterized by temporal bursty patterns.

To prove the burstiness of link creation, we consider for each user the event
time series where an event is represented by the creation of an edge incident
to the considered node. On each time series, we apply the technique proposed
in Vasquez \textit{et. al.}~\cite{Vazquez06} and extended by McGlohon
\text{et. al.}~\cite{McGlohon07}, both based on the inter-event time
distribution between consecutive events for a single individual.  If the edge
creation process is a Poisson-like process, \textit{i.e.} homogeneous, then
the inter-event time distribution should be an exponential distribution.  On
the other hand, a bursty arrival process is characterized by a power-law distribution
where many short time intervals, each corresponding to intensive activities
forming a burst, are separated by relatively fewer but longer periods of low
or zero activity.

\para{Results.} In order to distinguish if the process is homogeneous or
bursty, we fit the inter-event time data per node in our Renren dataset using
MLE (Maximum Likelihood Estimator), and select the model with the minimum AIC
(Akaike Information Criterion).  As a representative of the power law
distribution family, we choose the Pareto with exponential cutoff
$P(t)=t^{-\alpha}exp(-t/\lambda)$, and use the exponential distribution
$P(t)=\mu exp(-\mu)$ to describe the inter-event time Poisson process.
Finally, to avoid the impact of outliers, we remove from consideration users
who have too few events, \textit{i.e.} nodes with final degree less than 15
(median degree).

Our results show that minimum AIC is achieved by the Pareto distribution with
exponential cutoff, meaning almost all users in our dataset manifest a bursty
behavior in link creation.  In addition, the Kolmogorov-Smirnov (K-S test)
validates the selected hypothesis for almost all users ($86\%$ of the
population).  These measurements offer direct evidence that at the level of a
single individual, there is a heavy-tailed activity pattern.

Having shown that individuals add links in a temporal bursty manner, we
analyze the similarity of the bursty process across users, by computing the
distribution of the scale parameter $\alpha$ determined separately for each
user.  As shown in Figure~\ref{alphaDistro}, $\alpha$ values are scattered
around a peak at 1, with an heavy tail in the right side.  This partially
corroborates the results found in \cite{Vazquez06}, which showed a single
group of users with very similar behavior described by the Gaussian
distribution of $\alpha$ centered at 1.  However, the heavy tail suggests
that users in online social networks cannot be easily grouped in a single
category, but have quite different behaviors in adding links.

To understand the reasons behind the observed heavy tail, we take into
account two factors: the degree and the age of a node, \textit{i.e.} how long
the node has been in the network (in weeks).  In Figure~\ref{relations}, we show the
relationship between the scaling parameter $\alpha$ and the two variables we
consider.  Between $\alpha$ and the degree, we
observe that the mean $\alpha$ value increases with degree as shown
in Figure~\ref{alphaOnDegree}.  This fact suggests that nodes with higher
degree contribute more to the right tail.  This means that, although all the nodes
manifest the same bursty behavior, nodes with higher degree have more closely
spaced bursts.  With regards to age, shown in Figure~\ref{alphaOnAge} shows 
that age does not influence the right tail, since the mean value is
close to the mean of the $\alpha$ distribution, and remains quite constant for
different age values.  The small decrease is due to the fact that older nodes have
a greater chance to undergo long periods of inactivity.

In summary, we showed in this section that users follow a bursty process in creating
links, where bursts occur more frequently in nodes with high final degree.

\section{Degree Acceleration}
\label{sec4:inside}
Bursty phenomena have been studied in different areas of human activities,
such as clicks or queries in search engines~\cite{Parikh08}. 
However, these previous investigations focused on bursts resulting from
aggregate actions, such as group of users that manifest a
common interest at a certain time.  These burst detection algorithms are not suitable to
investigate per-node time sequences of link creation, or substructures inside bursts. 

\begin{figure}
\centering
\includegraphics[trim=0 0 -0.5in 0,width=3.5in]{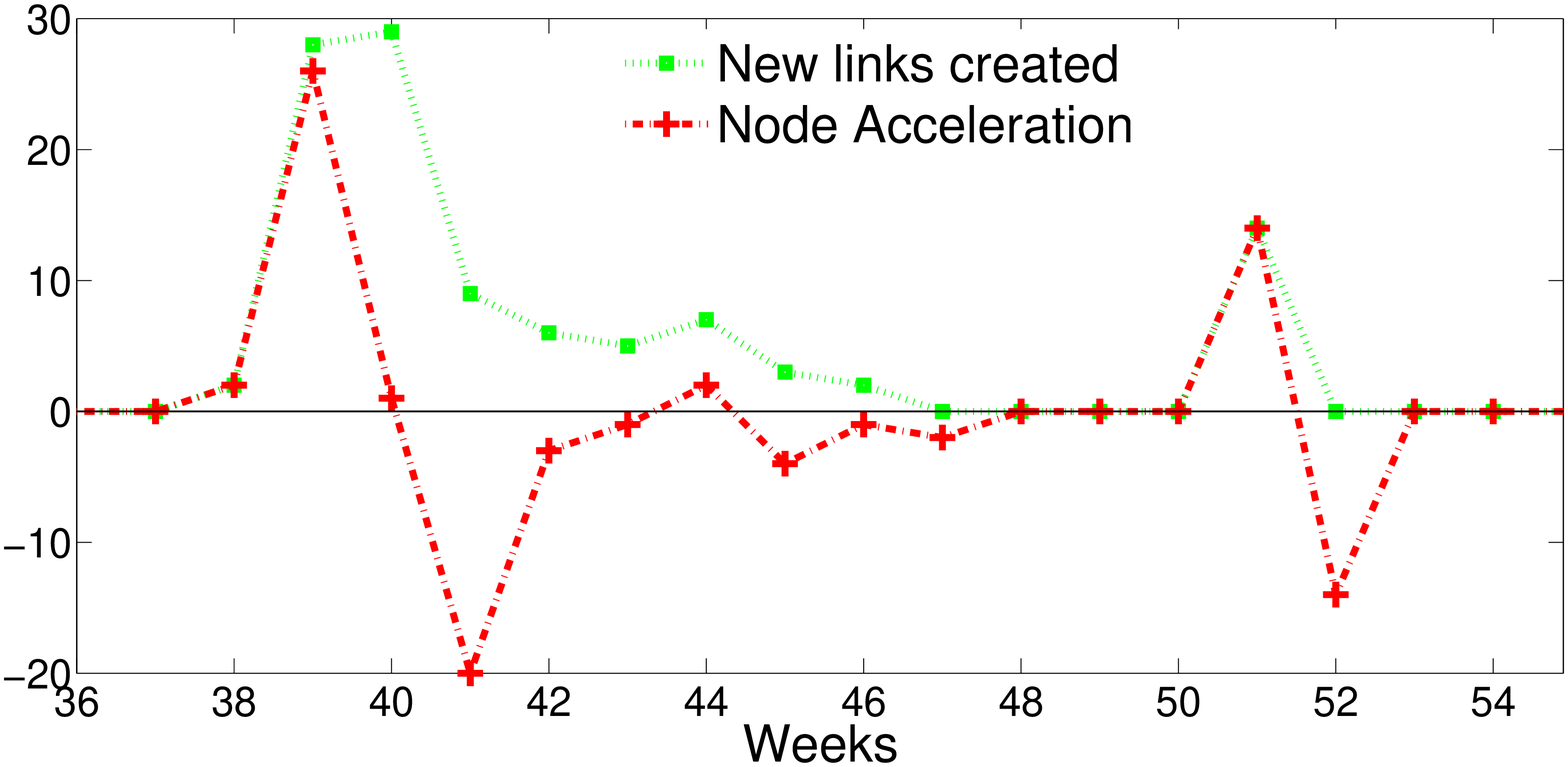}
\caption{An example of degree acceleration, computed on a single node from
  our dataset. The green dotted line represents the number of links created
  by the node each week. The red dotted line represents the acceleration
  computed according to Eq.~\ref{accDef}. In week 39, the node shows a large
  acceleration, follows by a plateau.  The node decelerates into week 42,
  when it enters a cruising phase (link creation is stable) for 4 weeks.}
\label{diagram}
\end{figure}

In this section, we propose a new methodology that identifies different
phases that make up the bursty nature of the link creation process, and
detects when bursts occur.  We also identify the role played by each phase
during the bursty process.  From a high level, we observe that the
alternation of activity/inactivity phases determines the burstiness of the
event trace.  In addition, bursts of activity have a typical internal
structure, composed by a rapidly increasing slope and a gradually decreasing
phase possibly interleaved by a plateau.  An example is shown in
Figure~\ref{diagram}.

\para{Degree Acceleration.}  Inspired by studies in physics and neuroscience on
highly dynamic systems~\cite{Chiappalone05}, we investigate the phases in
bursty processes and detect bursts by measuring significant increments and
decrements of new links formed per node.  A burst begins when link formation
activity rapidly increases, and ends following a decreasing phase.  By
leveraging the concept of acceleration, it is possible to easily identify and
quantify significant changes in link creation activity.  Let $d_{i}(t)$ be
the degree of node $i$ at time $t$, \textit{i.e.} the total number of links
incident to node $i$ at time $t$, and let $\Delta t$ be the time granularity
that interleaves each $d_{i}(t)$ measures.  We can then compute degree
acceleration as:

\begin{equation}\label{accDef}
a_{i}^{d}(t)=\frac{d_{i}(t)- 2d_{i}(t-\Delta t) + d_{i}(t-2\Delta t)}{(\Delta t)^{2} }
\end{equation}

By computing degree acceleration, we can observe the initial start of bursts
($a_{i}^{d}>> 0$) and a burst's decaying phase ($a_{i}^{d}<<0$).  An example
is shown in Figure~\ref{diagram}.  Note that acceleration captures
two types of steady state conditions:  a period of consistently high activity representing
the plateau inside an activity burst (after an initial acceleration phase), and a
steady state of low activity outside of activity bursts.

\para{Defining Phases.} While exploring the burstiness of the link creation
process, we found that the growth of each node is characterized by transition
phases in which users significantly change their link formation behavior. 
This led us to identify four different phases that describe the patterns
involved in the nodes' growth processes.  The different phases can be
described by defining a time-dependent state variable for each node in the
system.  More specifically, the acceleration phase is characterized by a
large increment in creating new links, i.e. $a_{i}^{d}>> 0$, and the
deceleration phase is described by a strong decay measured by $a_{i}^{d}<<0$.
Then we define two intermediate phases: {\em cruising} and {\em
  inactivity}. The first corresponds to a steady state of a node, where the
number of links created per week is almost constant.  This phase can
correspond both to high activity or to small oscillations around inactivity,
and is characterized by at least one new edge (captured by the variable
$c_{i}(t)=1$) and small $a_{i}^{d}$ values. These small $a_{i}^{d}$ values
are centered around the value $a_{i}^{d}=0$, and are bounded with two thresholds
$\theta_{1}$ and $\theta_{2}$.  The second phase, {\em i.e.} inactivity, occurs
when a node does not create any links for an entire time window.  We
formalize these four phases by introducing the function $s_{i}(t): \mathbb{R}
\rightarrow \left\{acc,dec,cruise, inact\right\}$ defined as follows:
\begin{equation}\label{eq:def}
s_{i}(t) = \left\{ 
  \begin{array}{l l}
    acc & \quad a_{i}^{d}(t) \in\left(\theta_{1},+\infty\right)\\
    dec & \quad a_{i}^{d}(t) \in\left(-\infty,\theta_{2}\right)\\
    cruise & \quad a_{i}^{d}(t)\in\left[\theta_{2},\theta_{1}\right]\wedge c_{i}(t)=1\\
    inact & \quad  c_{i}(t)=0\\
  \end{array} \right.
\end{equation}
where $c_{i}(t)=1$ if and only if node $i$ creates at least
one edge at time $t$, otherwise $c_{i}(t)=0$.  Degree acceleration
$a_{i}^{d}(t)$ and the related $s_{i}(t)$ function represent a general tool
to investigate the burstiness structure, and to highlight the
detailed properties of each phase. 

\section{Experimental Analysis}\label{sec4:phase}
In this section, we characterize the link creation process by analyzing our
Renren trace using our acceleration methodology.  The experimental analysis
has been performed with the following settings: $\Delta t=1$ week to avoid
cyclic fluctuations in acceleration due to increase in user activities over
each weekend, and cruising phase thresholds are $\theta_{1}=2$ and
$\theta_{2}=-2$.

\subsection{The Role of Phases}
The role played by each phase along the node lifetime is a key element to
understand the network dynamics, and is also crucial when designing 
generative models based on per-node temporal behavior. To this purpose, we
consider two main aspects: (i) the time a node spends in each phase and (ii)
the per-node amount of links created in the different phases.

We perform this analysis from two perspectives, by considering the aggregate
behavior of all nodes, and on per-node behavior.  In order to
understand the role of different phases during a node's lifetime, we define
$\phi^{l}$ and $\psi^{l}(i)$ to compute the percentage of time spent in each
phase by all nodes (Equation~\ref{phiLife}) and by each node
(Equation~\ref{psiLife}).
\begin{equation}\label{phiLife}
\phi_{phase}^{l}=\dfrac{\sum_{i\in N}\sum_{t=1}^{T}\mathcal{I}_{phase}(s_{i}(t))}{\sum_{i\in N}life(i)}
\end{equation}
\begin{equation}\label{psiLife}
\psi^{l}_{phase}(i)=\dfrac{\sum_{t=1}^{T}\mathcal{I}_{phase}(s_{i}(t))}{life(i)}
\end{equation}
where $life(i)$ represents the lifetime in weeks of a node,
$\mathcal{I}$ is the indicator function and
$phase=\left\{acc,dec,cruise,inact\right\}$. $N$ indicates the number of
nodes at time $T$, which represents the last week considered in the
dataset. 

The relationship between link creation and phase is quantified by $\phi^{e}$, which corresponds to the
percentage of the overall edges created within each phase, and $\psi^{e}(i)$, which is the link generation rate for node $i$ in each phase:
\begin{equation}\label{phiEdge}
\phi_{phase}^{e}=\dfrac{\sum_{i\in N}\sum_{t=1}^{T}\mathcal{I}_{phase}(s_{i}(t))n_{i}(t)}{2m}
\end{equation}
\begin{equation}\label{psiEdge}
\psi^{e}_{phase}(i)=\dfrac{\sum_{t=1}^{T}\mathcal{I}_{phase}(s_{i}(t))n_{i}(t)}{d_{i}(T)}
\end{equation}
where $m$ is the number of link at time T. The results are
reported in Table \ref{tab:phases}, where $\psi^{l}_{0.8}$ and
$\psi^{e}_{0.8}$ are the $0.8$-quantile of the distributions of $\psi^{l}$
and $\psi^{e}$, and are discussed below. 
\begin{table}
\centering
    \begin{tabular}{ | c | c | c | c | c |}
    \hline
     & acc & dec & cruise & inact \\ \hline
    $\phi^{l}$ & 0.11 & 0.14 & 0.28 & 0.47 \\ \hline
    $\phi^{e}$ & 0.52 & 0.17 & 0.31 & 0 \\ \hline
    $\psi^{l}_{0.8}$ & 0.25 & 0.27 & 0.44 & 0.66 \\ \hline
    $\psi^{e}_{0.8}$ & 0.74 & 0.70 & 0.16 & 0 \\ 
    \hline
    \end{tabular}
\caption{ In the first two rows we report $\phi^{e}$ (definition \ref{phiEdge}) and $\phi^{l}$ (definition \ref{phiLife})  values for each phase. In the last two rows the $0.8$-quantiles of $\psi^{l}$ and $\psi^{e}$ distributions.}
\label{tab:phases}
\end{table}

\para{Inactivity phase.} During the inactivity phase, by definition, we do
not observe growth since no links are created. However, inactivity acquires
importance in the temporal dimension, because it deeply affects the
burstiness. The high values of $\phi^{l}$ and $\psi^{l}_{0.8}$ highlight that
node activities are concentrated in few and small periods; thus, for most
part of their life, nodes do not influence the network dynamic evolution.


\para{Acceleration and Deceleration Phases.} Nodes spend only a small amount
of their life in these phases, in particular after acceleration events,
longer period of weaker activity follow. However, the amount of links
generated in these phases determines the structure of Renren. In fact, a link
has very high probability, $69\%$, to be generated in one of these two
phases, in particular $52\%$ in acceleration and $17\%$ in deceleration.

\begin{figure}
\centering
\includegraphics[trim=0 0 -0.5in 0,width=3.5in]{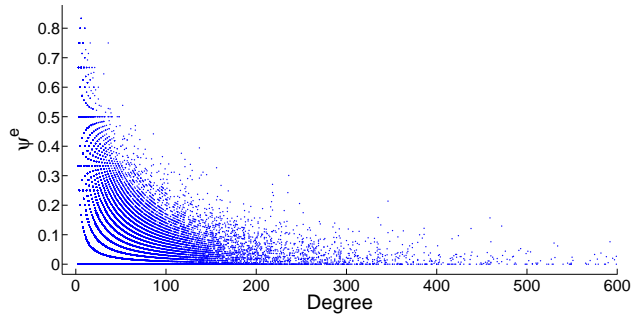}
\caption{Relationship between $\psi^{e}_{cruise}$ and the
  degree. $\psi^{e}_{cruise}$ decreases as the degree raises, so the cruising
  phase has a pronounced impact only for nodes with low degree.}
\label{cruisingVSdegree}
\end{figure}

\para{Cruising Phase.}  Cruising periods cover an important portion of nodes'
lifetime. Furthermore, $\phi^{e}=0.31$ and $\psi^{e}=0.16$ would suggest that
this phase has a role also in link creation. However, only few cruising
periods have relevance in the edge growth. Indeed, it depends on whether the
cruising phase is inside a burst or it corresponds to small oscillations
around inactivity. A node in a burst cruising phase is creating many links,
while in the other case the number of links created is irrelevant.  Finally,
the cruising phase has a pronounced impact only for nodes with low degree, as
shown in Figure \ref{cruisingVSdegree}.

We have shown that acceleration ($acc$) and deceleration ($dec$) phases are
those responsible of the growth and dynamics of the network, despite the fact
that they represent a very small part of a node life.

\subsection{Acceleration and Deceleration Features}
\label{sec4:accDec}
\begin{figure}[t]
\centering
\subfigure[]{\includegraphics[trim=31pt 0pt 7pt 0pt,width=4.1cm]{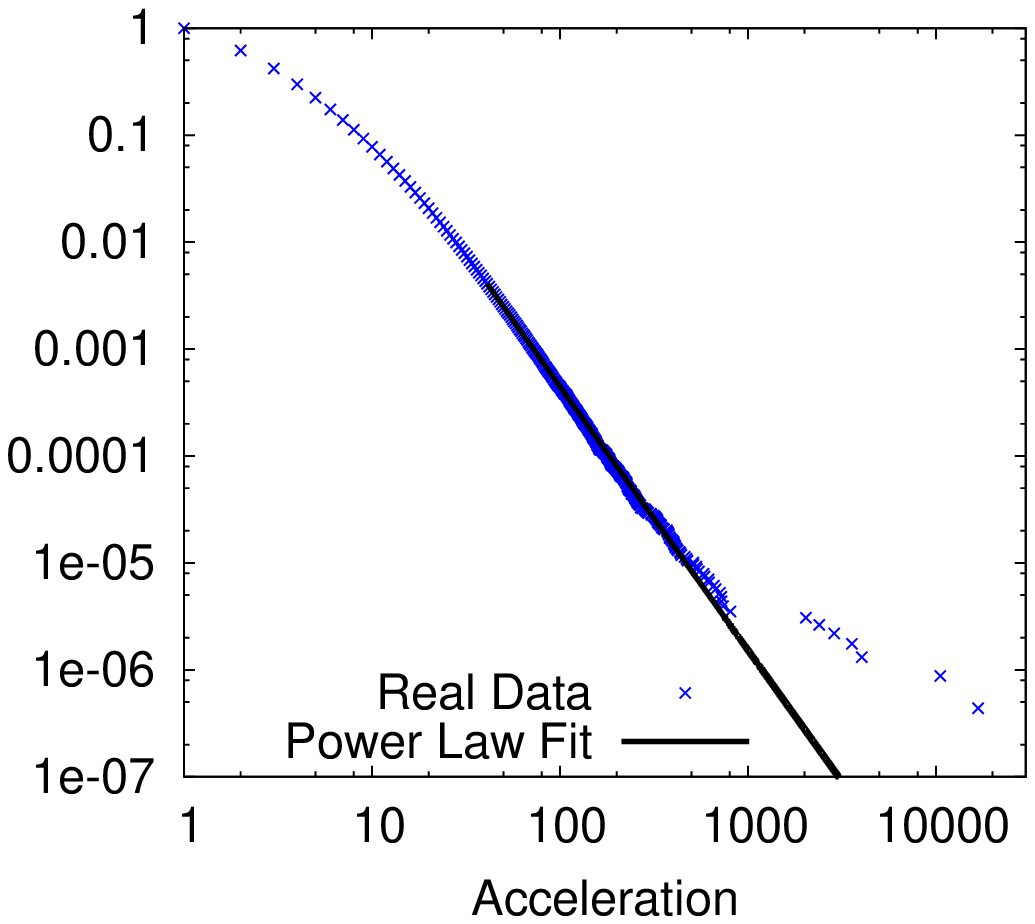}
\label{AccCCDF}}
\subfigure[]{\includegraphics[trim=31pt 0pt 7pt 0pt,width=4.1cm]{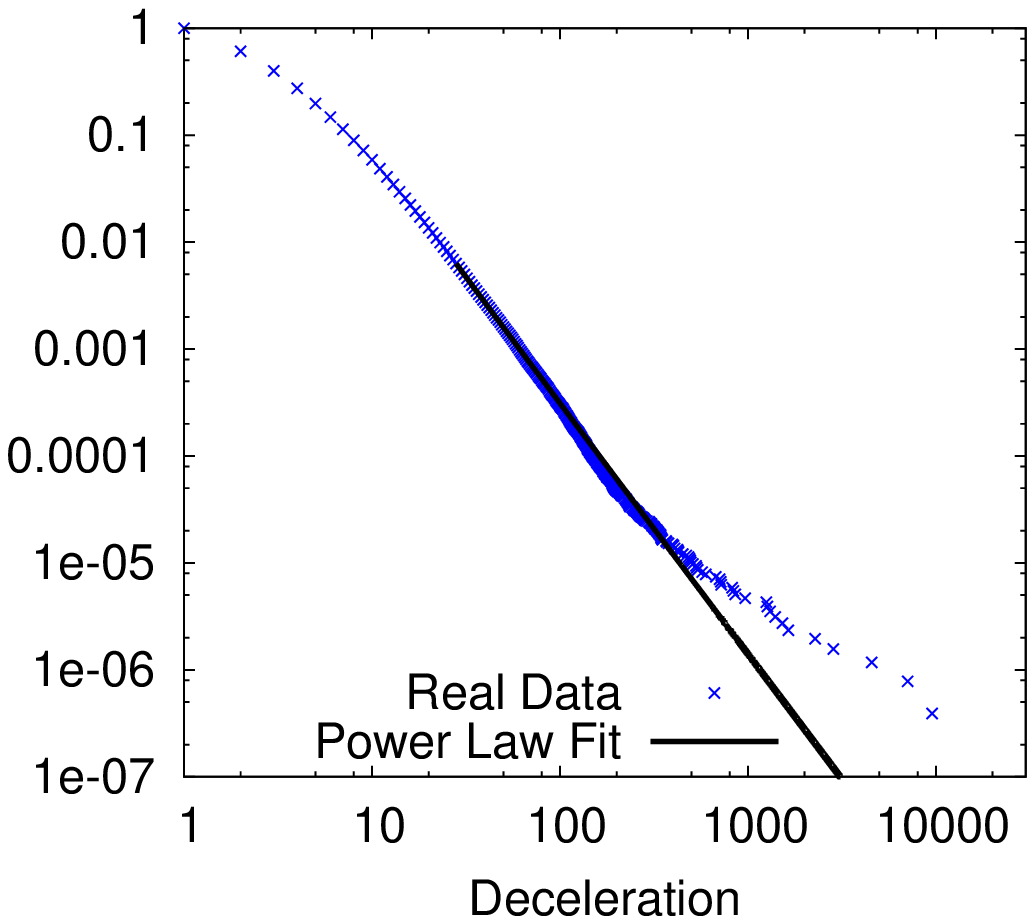}
\label{DecCCDF}}
\caption{ a) Acceleration CCDF and the resulting fitted distribution
  ($\alpha=3.46$). b) Deceleration CCDF and the resulting fitted distribution
  ($\alpha=3.34$).}
\label{aging}
\end{figure}

In depth understanding of acceleration/deceleration phases reveals how users
operate in the network after they join. This knowledge could be very useful
to ensure efficient management of the OSN's resources. This section focuses
on acceleration and deceleration by means of illustrating their importance
from a network perspective; showing that they follow a power law distribution
and finally investigating the impact of node aging on link creation process. 

\begin{figure}[t]
\centering
\includegraphics[trim=0 0 -0.5in 0,width=3.5in]{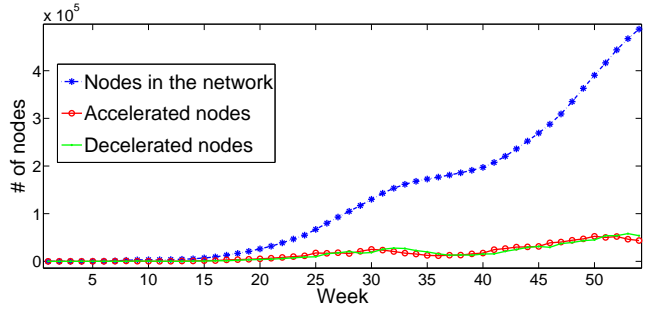}
\caption{For each week, the number of nodes in the network (network size) and the number of nodes in the acceleration/deceleration phases. In each week only a very small number of nodes
are in the acc/dec phases though the network size rapidly grows.}
\label{overall}
\end{figure}

\para{Network perspective on acceleration/deceleration.} From the network
perspective, an estimate of how many and which nodes are changing the graph
structure would greatly help in managing the system resources. As it can be
seen in Figure \ref{overall}, in each week only a very small number of nodes
are acc/dec phases; for example at the end of the year they are almost
20\%. These nodes can be easily identified as soon as they experience a phase
transition from the inactive/cruising to the accelerated phase since their
values of acceleration abruptly increase. 
\begin{figure}[t]
\centering
\subfigure[]{\includegraphics[width=3.5in]{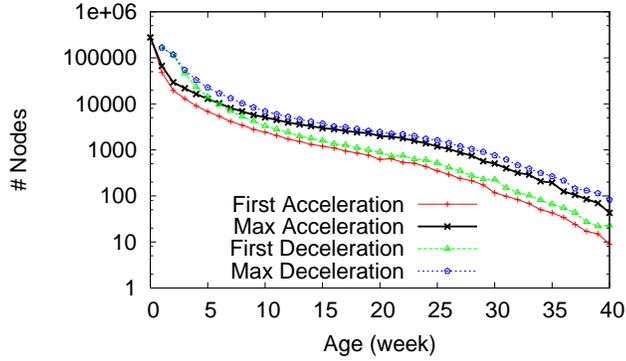}
\label{firstMaxBurst}}
\subfigure[]{\includegraphics[width=3.5in]{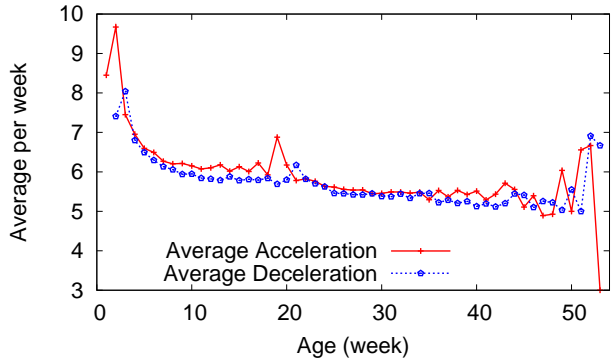}
\label{ageAndAcc}}
\caption{ a) shows the times when nodes first
  experience their first acceleration, maximum acceleration, first
  deceleration, and maximum deceleration ($y$-axis on logscale). b) shows the
  average acceleration/deceleration with respect to node age.}
\label{aging2}
\end{figure}

\para{Acceleration/deceleration probability distributions.}  By applying the
statistical framework proposed by Clauset~\cite{Clauset09}, we find that
acceleration and deceleration distributions are power law,
(Figure~\ref{AccCCDF} and~\ref{DecCCDF}). By considering the overall network,
this result implies that half of the acceleration and deceleration events have a
small size, but they are very likely to show rapid increase and decrease respectively. The
upper tail of the acceleration distribution exhibits so high values of
acceleration that can't correspond to normal user. Those events are most
likely associated to people with a large amount of followers or accounts for
advertisement.

\para{The impact of aging.} The general behavior of a node is a sequence of acceleration/deceleration phases of 
constant magnitude, after an initial burst.  In general, nodes wait
at most for one month before initiating their activity.

We start by defining the $age(t)$ of a node $u$ as the time elapsed between
the appearance of $u$ in the network (timestamp of the first edge incident to
$u$), and time $t$.  The observables whose dependence on age, need to be
studied are: $n_{firstAcc/Dec}(t)$, the number of nodes showing their first
acceleration / deceleration at time $t$ and $n_{maxAcc/Dec}(t)$, the number of
nodes manifesting their maximum acceleration / deceleration at time
$t$. Finally, we calculate the
average acceleration/ deceleration $avg_{Acc/Dec}(age)$.

Analyzing and comparing $n_{maxAcc}(t)$ and $avg_{Acc}(age)$ in
Figures~\ref{firstMaxBurst} and \ref{ageAndAcc}, we observe that most nodes
enter the phase of maximum acceleration in the first week. In addition,
Figure~\ref{firstMaxBurst} shows that the activity after the first peak does
not decrease as fast as its respective acceleration.

Figure~\ref{ageAndAcc} highlights another interesting behavior
of the $acc/dec$ phases. The average acceleration remains
constant when age increases. This is consistent with what we found in 
Figure~\ref{diagram}, {\em i.e.} nodes experience a big burst of acceleration
in the first week after joining the network, and subsequent bursts never
match the first in intensity.



\section{Conclusion}
In this paper we investigated the bursty nature of the link creation process
in OSN. We prove not only that it is an highly inhomogeneous process, but
also identify patterns of burstiness common to all nodes.  In terms of edge
creation, users are inactive for most of their lifetimes, and concentrate
their link activity in a number of short regular time periods. To
characterize node activity, we develop a new methodology based on the
acceleration of degree growth, which allows us to highlight the internal
structure of link creation bursts.

We believe using acceleration as a general metric to characterize network
dynamics prompts future work in studying link generation mechanisms. In
particular, defining different phases of edge creation hints at the
possibility of characterizing users into distinctive activity levels that
correlate with their likelihood of adding social links.  Some preliminary
results confirm this intuition: when nodes (users) first join the network,
they create links based on the preferential attachment mechanism; while in
later bursts, nodes seem to explore (acceleration phase) and densify (deceleration)
in far regions of the graph. These results open the door for new generative models
that consider different phases of node activity.  

\balance

\bibliographystyle{splncs}
\bibliography{ale,han,zhao,matteo}

\begin{thebibliography}{10}

\bibitem{leskovec2005graphs}
Leskovec, J., Kleinberg, J., Faloutsos, C.:
\newblock Graphs over time: densification laws, shrinking diameters and
  possible explanations.
\newblock In: Proc. of {KDD}. (2005)

\bibitem{Vazquez06}
V\'azquez, A., Oliveira, J.a.G., Dezs\"o, Z., Goh, K.I., Kondor, I.,
  Barab\'asi, A.L.:
\newblock Modeling bursts and heavy tails in human dynamics.
\newblock Physical Review E \textbf{73} (2006)

\bibitem{Kleinberg_burst}
Kleinberg, J.:
\newblock Bursty and hierarchical structure in streams.
\newblock In: Proc. of KDD. (2002)

\bibitem{renren-imc10}
Jiang, J., Wilson, C., Wang, X., Huang, P., Sha, W., Dai, Y., Zhao, B.Y.:
\newblock Understanding latent interactions in online social networks.
\newblock In: Proc. of {IMC}. (2010)

\bibitem{Palla2007}
Palla, G., Barabasi, A., Vicsek, T.:
\newblock {Quantifying social group evolution}.
\newblock Nature \textbf{446} (2007)  664--667

\bibitem{Berlingherio2010}
Berlingerio, M.,  et~al.:
\newblock As time goes by: Discovering eras in evolving social networks.
\newblock In: Proc. of PAKDD. (2010)  81--90

\bibitem{mcglohon2008weighted}
McGlohon, M., Akoglu, L., Faloutsos, C.:
\newblock Weighted graphs and disconnected components: patterns and a
  generator.
\newblock In: Proc. of {KDD}. (2008)

\bibitem{zheleva2009co}
Zheleva, E., Sharara, H., Getoor, L.:
\newblock Co-evolution of social and affiliation networks.
\newblock In: Proc. of {KDD}. (2009)

\bibitem{akoglu2009rtg}
Akoglu, L., Faloutsos, C.:
\newblock Rtg: a recursive realistic graph generator using random typing.
\newblock Machine Learning and Knowledge Discovery in Databases (2009)  13--28

\bibitem{Backstrom2006}
Backstrom, L., Huttenlocher, D., Kleinberg, J., Lan, X.:
\newblock Group formation in large social networks: membership, growth, and
  evolution.
\newblock In: Proc. of KDD. (2006)

\bibitem{leskovec2008microscopic}
Leskovec, J., Backstrom, L., Kumar, R., Tomkins, A.:
\newblock Microscopic evolution of social networks.
\newblock In: Proc. of {KDD}. (2008)

\bibitem{dorogovtsev}
Dorogovtsev, S.N., Mendes, J.F.F.:
\newblock Accelerated growth of networks.
\newblock CoRR (2002)

\bibitem{barabasi:nature2005}
Barab\'asi, A.L.:
\newblock The origin of bursts and heavy tails in human dynamics.
\newblock Nature \textbf{435} (2005)  207--211

\bibitem{McGlohon07}
McGlohon, M.,  et~al.:
\newblock Finding patterns in blog shapes and blog evolution.
\newblock In: Proc. of ICWSM. (2007)

\bibitem{kumar11}
Jo, H.H., Pan, R.K., Kaski, K.:
\newblock Emergence of bursts and communities in evolving weighted networks.
\newblock PLoS ONE \textbf{6} (08 2011)  e22687

\bibitem{Eom11}
Eom, Y.H., Fortunato, S.:
\newblock Characterizing and modeling citation dynamics.
\newblock PLoS ONE \textbf{6}(9) (2011)

\bibitem{Crovella97}
Crovella, M.E., Bestavros, A.:
\newblock Self-similarity in world wide web traffic: evidence and possible
  causes.
\newblock IEEE/ACM Transactions on Networking \textbf{5} (1997)  835--846

\bibitem{Parikh08}
Parikh, N., Sundaresan, N.:
\newblock Scalable and near real-time burst detection from ecommerce queries.
\newblock In: Proc. of KDD. (2008)

\bibitem{Chiappalone05}
Chiappalone, M.,  et~al.:
\newblock Burst detection algorithms for the analysis of spatio-temporal
  patterns in cortical networks of neurons.
\newblock Neurocomputing \textbf{65-66} (2005)  653--662

\bibitem{Clauset09}
Clauset, A., Shalizi, C.R., Newman, M.E.J.:
\newblock Power-law distributions in empirical data.
\newblock SIAM Review \textbf{51} (2009)  661--703

\end{thebibliography}

\end{document}